\documentclass{rjparticle}
\usepackage{graphicx}

\newcommand{\miktex}{\hbox{Mik\kern-.15em\TeX}}

\newcommand{\be}{\begin{equation}}
\newcommand{\ee}{\end{equation}}
\newcommand{\bea}{\begin{eqnarray}}
\newcommand{\eea}{\end{eqnarray}}

\title{Vanishing dimensions: theory and phenomenology
}
\author[1]{Dejan Stojkovic}
\affil[1]{HEPCOS, Department of Physics,
SUNY at Buffalo, Buffalo, NY 14260-1500\\Email:{\em ds77@buffalo.edu}}

\begin{document}
\maketitle
\begin{abstract}
Lower-dimensionality at higher energies has manifold theoretical advantages as recently pointed out. Moreover, it appears that experimental evidence may already exists for it - a statistically significant planar alignment of events with energies higher than TeV has been observed in some earlier cosmic ray experiments. If this alignment is not a fluke, then the LHC should be able to see effects associated with the dimensional crossover. Further, (2+1)-dimensional spacetimes have no gravitational degrees of freedom, and gravity waves cannot be produced in that epoch in the early universe. This places a universal maximum frequency at which primordial gravity waves can propagate, which may be accessible to future gravitational wave detectors such as LISA. In this talk, the theoretical motivation for "vanishing dimensions" as well as generic experimental and observational signature will be discussed.
\end{abstract}

{\it Introduction.}~
Despite the fascinating success of the modern physics embodied in the Standard Model of particle physics and the Standard Model of Cosmology, serious problems have accumulated that need prompt attention. It is very difficult to imagine further model building in both particle physics and cosmology without addressing fundamental problems like the Standard Model hierarchy problem, the problem of initial conditions in cosmology, the cosmological constant problem, the dark matter problem, the black hole information loss problem etc. It is becoming increasingly clear that straightforward extensions of existing theories will not suffice and some radically new ideas are needed. Here, we propose a new approach. Instead of changing the mathematical details of the well established physical theories, we can change the background on which these theories are formulated. This shift will create enough room for looking at the many fundamental problems from a completely new angle.

\begin{figure}[h]
\scalebox{0.25}{\includegraphics{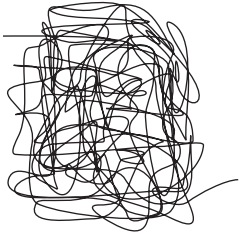}}
\caption{An example of a structure which is one-dimensional on short scales while it appears effectively two-dimensional at large scales.}
\label{1d2d} \end{figure}

Most of the fundamental problems that we are currently facing stem from the ultraviolet (short distance) and infrared (large distance) divergencies. We believe that (with some exceptions) we understand our universe on scales approximately between $10^{-17}$ cm and a Gpc. The first scale corresponds to the energy scale of TeV$^{-1}$ which is the energy probed in the highest energy accelerators available so far. The second scale corresponds to the distance characteristic for super-clusters of galaxies, i.e. the scale at which cosmology kicks in. At scales shorter than $10^{-17}$ cm and larger than about a Gpc we are running into problems.  At the intermediate scales between $10^{-17}$ cm and a Gpc we know pretty well that our space is three dimensional. We propose that the space at scales shorter than $10^{-17}$ cm is lower dimensional, while at scales larger than a Gpc the space is higher dimensional.
The most natural setup in this sense would be the one in which the effective number of dimensions increases with the length scale. At the fundamental level, the space-time is probably quantized, i.e. discrete. We leave aside the problem of time and consider only the evolution of space. At the shortest distances at which our space appears as continuum, the space is one-dimensional. At a certain critical length scale, the space becomes effectively two-dimensional (see Fig.~\ref{1d2d}). At the scale of about $10^{-17}$ cm, the space becomes effectively three-dimensional. Finally, at the scales of about a Gpc, the space becomes effectively four-dimensional. In principle, this hierarchy does not need to stop a priori at any finite number. However, it would be interesting for other reasons if this construct stops at $10$ or $26$ effective dimensions. In a dynamical picture where the universe starts from zero size and then grows, there is no background with the fixed number of dimensions in which the universe expands. The expanding structure actually builds the effective dimensions during its evolution.


An evolving background with such features can be effectively described with a formalism of an induced metric on a sub-manifold  embedded in a higher dimensional manifold. The induced metric can be written as $\gamma_{ab} = g_{\mu \nu} \partial_a X^\mu \partial_b X^\nu$ where $g_{\mu \nu}$ is the background metric on a higher dimensional manifold, Latin indices cover the induced metric, while Greek indices cover the background metric. Usually, $g_{\mu \nu}$ is considered to be the fundamental metric of the space time, while $\gamma_{ab}$ is the induced one on the sub-manifold. In our approach, the roles are inverted. $\gamma_{ab}$ is the fundamental metric since it builds the effective foreground (rather than background) metric $g_{\mu \nu}$. The model formulated in \cite{Karliner:1988hd} (with quite a different motivation) can be nicely used for this purpose.
In that model, a string with the manifest tendency to grow was considered. The string was chopped into $N$ segments, each carrying longitudinal momentum $P^+/N$, where $P^+$ is the total longitudinal momentum. The Hamiltonian for such a string is given by
\be
H = \sum_i^N \frac{1}{2P^+(i)}\{ \vec{P}^2_\perp (i) + \left[ \vec{X}(i+1) - \vec{X}(i) \right]^2 \}
\ee
where $\vec{P}_\perp (i)$ and  $\vec{X}(i)$ are the transverse momentum and position of the i-th segment. The non-vanishing longitudinal momentum causes the string the grow in length, while the non-vanishing transverse momentum is the source of the string's extrinsic curvature. Thus, such a string builds a structure described in Fig.~\ref{1d2d}. It can be shown \cite{Karliner:1988hd} that the total length of the string grows as $N$, while the radius of the effective space covered by such a string grows as $\sqrt{\log N}$. In the limit of $N \rightarrow \infty$ the string becomes space-filling.

{\it Curing ultraviolet divergences.}~
There exist a strong motivation for changing the dimensionality of the space-time at small distances. One of the most acute problems connected with ultraviolet divergences is the Standard Model hierarchy problem.  The Higgs Lagrangian together with Yukawa couplings in the Standard Model is
\begin{equation}
L_H=D_\nu \Phi ^\dagger D^\nu \Phi - \mu^2 \Phi^\dagger \Phi + \frac{\lambda}{2}(\Phi^\dagger \Phi)^2 - \sum_f g_f \Phi \bar{\psi_f} \psi_f
\end{equation}
where $\Phi$ is the Higgs field, $g_f^2=m_f^2 /v^2$,  $\lambda = m_H^2/(2v^2)$, $m_f$ and $m_H$ are the fermion and Higgs mass respectively, while $v$ is the Higgs field vacuum expectation value.
If we consider all 1 loop one-particle-irreducible diagrams, then we find that the Higgs self energy comes from the three types of diagrams. In $3+1$ dimensions, all of these terms are quadratically divergent with the cut-off scale $\Lambda$ at which new physics appears. The contribution of fermions, gauge bosons and the Higgs itself in the loop are respectively
\begin{eqnarray} \label{terms}
&&  i \frac{g_f^2}{2}\int^\Lambda \frac{d^4k}{(2\pi)^4} tr(\frac{i}{\not k -m_f}\frac{i}{\not k+\not{p} -m_f})   \sim -\Lambda^2 \, \frac{g_f^2}{32\pi^2}  \nonumber \\
&&  \ \ \ \ \ \ \ i \frac{g^2}{4} \int^\Lambda \frac{d^4k}{(2\pi)^4} \frac{1}{k^2-m_g^2} \ \sim \ \Lambda^2 \, \frac{g^2}{64\pi^2} \\
&& \ \ \ \ \ \ \
i 3\lambda \int^\Lambda \frac{d^4k}{(2\pi)^4} \frac{1}{k^2-m_H^2} \ \sim \ \Lambda^2 \, \frac{3\lambda}{16\pi^2} \nonumber
\end{eqnarray}
Here, $g$ is the gauge coupling constant, $g^2=2m_g^2/v^2$, while
$m_g$ is the mass of gauge boson.
The quadratic divergence implies that a very special cancelation needs to happen between the bare Higgs mass and the corrections, unless the scale of new physics $\Lambda$ is very close to the electroweak scale.

Solutions to the hierarchy problem proposed so far postulate new physics not very far from the electroweak scale.
An alternative approach would be to keep the Standard Model physics, and change the dimensionality of the background on which the model is defined. For example, in $2+1$ dimensional space-time all of the terms in (\ref{terms}) are only linearly divergent

\begin{eqnarray}
&& i \frac{g_f^2}{2}\int^\Lambda \frac{d^3k}{(2\pi)^3} tr(\frac{i}{\not k -m_f}\frac{i}{\not k+\not{p} -m_f}) \ \sim \ -\Lambda \, \frac{g_f^2}{4\pi^2}  \nonumber \\
&&  \frac{g^2}{4} \int^\Lambda \frac{d^3k}{(2\pi)^3} \frac{1}{k^2-m_g^2} \ \sim \  \Lambda \, \frac{g^2}{8\pi^2}  \\
&&  i 3\lambda \int^\Lambda \frac{d^3k}{(2\pi)^3} \frac{1}{k^2-m_H^2} \ \sim \  \Lambda \, \frac{3\lambda}{2\pi^2}  \nonumber
\end{eqnarray}

Going further, in $1+1$ dimensional space-time all of these terms are only logarithmically divergent

\begin{eqnarray}
&&  i \frac{g_f^2}{2}\int^\Lambda \frac{d^2k}{(2\pi)^2} tr(\frac{i}{\not k -m_f}\frac{i}{\not k+\not{p} -m_f}) \ \sim \ -\log(\Lambda/m_f) \, \frac{g_f^2}{4\pi}   \nonumber\\
&&  \frac{g^2}{4} \int^\Lambda \frac{d^2k}{(2\pi)^2} \frac{1}{k^2-m_g^2} \ \sim \ \log(\Lambda/m_g) \, \frac{g^2}{8\pi} \\
&&  i 3\lambda \int^\Lambda \frac{d^2k}{(2\pi)^2} \frac{1}{k^2-m_H^2} \ \sim \ \log(\Lambda/m_H) \, \frac{3\lambda}{2\pi}  \nonumber
\end{eqnarray}

Thus, keeping the Standard Model Lagrangian and lowering the dimensionality of the space-time greatly improves the fine tuning problem in the Standard Model. In fact the dimensional regularization procedure tells us that the ultraviolet divergences in field theory are poles in the dimension plane. Lowering the dimensionality of the space-time universally cures ultraviolet divergences in practically all of the field theories.

Possible implications of having less dimensions at higher energies are very important for the LHC physics.  There are three immediate and spectacular consequences of this model at the LHC, which should be observable if the dimensional crossover scale is  $\sim 1$~TeV, {\it i.e.\/}, within the reach of the machine: {\it (i)\/} cross section of hard scattering processes changes compared to that in the SM as the $Q^2$ becomes comparable with the crossover scale; {\it (ii)\/} $2 \to 4$ and higher order scattering processes at high energies become planar, resulting, {\it e.g.\/}, in four-jet events, where all jets are produced in one plane in their center-of-mass frame, thus strikingly different from standard QCD multijet events; {\it (iii)\/} under certain conditions, jets of sufficiently high energy may become elliptic in shape (for details see \cite{Anchordoqui:2010hi,Anchordoqui:2010er}).

What about gravity? The most elusive concept in modern physics - quantum gravity - is much more within the reach in lower dimensions. Gravity in $3+1$ dimensions is complicated, highly nonlinear, perturbatively non-renormalizable theory. All of the attempts to quantize gravity in $3+1$ dimensions failed so far. However, we realize that if the fundamental short scale physics is lower dimensional, there is no need to quantize $3+1$ dimensional gravity. Instead we should quantize  $2+1$ and $1+1$ dimensional gravity. This is much easier task to accomplish. In any space-time, the curvature tensor $R_{\mu \nu \rho \sigma }$ may be decomposed into a Ricci scalar $R$, Ricci tensor $R_{\mu \nu }$ and conformally invariant Weyl tensor $C_{\mu \nu \rho }^\sigma $. In $2+1$ dimensions the Weyl tensor vanishes and  $R_{\mu \nu \rho \sigma }$ can be expressed solely through $R_{\mu \nu }$ and $R$.  Explicitly
\be
R_{\mu \nu \rho \sigma } = \epsilon_{\mu \nu \alpha} \epsilon_{\rho \sigma \beta} G^{\alpha \beta}
\ee
This in turn implies that any solution of the vacuum Einstein's equations is locally flat. Thus, $2+1$  dimensional space-time has no local gravitational degrees of freedom, i.e. no gravitational waves in classical theory and no gravitons in quantum theory. The number of degrees of freedom in such a theory is finite, quantum field theory reduces to quantum mechanics and the problem of non-renormalizability disappears \cite{Carlip:1995zj}. Obviously, $2+1$ dimensional gravity has much nicer structure than its $3+1$ dimensional cousin. For the reason of simplicity, $1+1$ dimensional gravity is even more attractive. Einstein's action in $1+1$  dimensional space-time is a constant (Euler's characteristic of the manifold in question) and the theory is trivial (unless augmented by some additional fields). Models of gravity in $1+1$ dimensions are completely solvable \cite{Klosch:1997md,LouisMartinez:1993cc} and a lot of work has been done on their quantum aspects \cite{Grumiller:2006ja,Zaslavskii:2003eu,Giddings:1992ae,Callan:1992rs,Bogojevic:1998ma}.

If the space-time is $2+1$ dimensional at distances shorter than $10^{-17}$ cm, then we expect some strong implications for the high energy scattering processes.
The structure of well established theories like Quantum Chromo Dynamics (QCD)  becomes much simpler in $2+1$ and $1+1$ dimensional space-times.The form of the QCD Lagrangian in $2+1$ dimensions is the same as in  $3+1$ dimensions
\begin{eqnarray}
&& {\cal L} = -\frac{1}{4}F^a_{\mu \nu } F^{a\mu \nu } + i \bar{\psi}\gamma^\mu (\partial_\mu + igA^a_\mu T^a)\psi  \\
&& F^a_{\mu \nu} = \partial_\mu A^a_\nu - \partial_\nu A^a_\mu + g f^{abc} A_\mu^b A_\nu ^b \, , \nonumber
\end{eqnarray}
except that the $\gamma$-matrices can be chosen to be proportional to two-dimensional Pauli matrices, while spinors $\psi$ are two-component spinors.
It is interesting that $2+1$ dimensional QCD is super-renormalizable, i.e. only a finite set of graphs need overall counter terms. This is a consequence of the fact that the coupling constant in this theory has positive dimension.
In  $2+1$ dimensional space-time there is only one transverse dimension, so there is no arbitrarily high transverse angular momentum. This implies that there exist no Regge-like behavior due to  exchange of states of high spin which is characteristic in $3+1$ dimensional space-time. For the LHC, it is certainly very important to calculate hadron-hadron scattering amplitude. It is not difficult to verify that the result is quite different from the standard one. The total cross section falls off like $1/\log s$, where $s$ is the center of mass energy squared. Characteristic Regge factor $s^\alpha, \alpha >0$ is completely absent \cite{Li:1994et}, in the strong contrast with $3+1$ dimensions.

For completeness, we mention $1+1$ dimensional QCD.
QCD in  $1+1$ dimensions is trivially asymptotically free, being super-renormalizable. The model on the infinite line has no gluon degrees of freedom, but it is a self-interacting fermion theory. In QCD on a circle, boundary conditions force the retention of quantum mechanical (zero-mode) gauge degrees of freedom in a Hamiltonian formulation. The dynamics of the zero modes lead to an elimination of fermionic non-singlet states from the spectrum in the continuum limit \cite{Engelhardt:1995qm}, thus practically eliminating color. While the details of $1+1$ and especially  $2+1$ dimensional QCD are very interesting for collider phenomenology, they  can not help (at least not directly) with the problem of quark confinement in a nucleon since the effective size of a nucleon (GeV$^{-1}$) is much larger than the critical distance at which the space appears $3+1$ dimensional (TeV$^{-1}$).

{\it Curing infrared divergences.}~
From the dimensional regularization point of view, infrared divergences, like their cousins ultraviolet ones, are also poles in the dimension plane. Therefore, changing the effective dimensionality of the space-time will remove them trivially.
For example, the electron-photon interaction cross section is divergent at low momenta (large wavelengths) of the virtual photon. The probability amplitude P for this process (see e.g. \cite{PS}) is
\begin{equation}
{\rm P} \approx \frac{\alpha}{\pi}\int^{|q|}_0 dk \frac{1}{k}I(v,v')
\end{equation}
Here, $k$ is the momentum of the virtual photon, $|q|$ is its  maximal value, while $v$ and $v'$ are initial and final velocity of the electron. $I(v,v')=dE/dk$ is the intensity, which is independent of $k$ when $k$ is small. The probability is obviously divergent at low $k$. We can remove this divergence by introducing an infrared momentum cutoff at very low energies. This can be ad-hock justified as a limitation of a finite size detector. However, changing the effective dimensionality of the space-time at large scales also removes this divergence. In our context, if the space-time becomes effectively $4+1$ dimensional the integral becomes
\begin{equation}
{\rm P} \propto \int^{|q|}_0 dk
\end{equation}
This integral is not divergent as $k\rightarrow 0$ and the infrared divergence is cured. Thus, increasing the dimensionality of the space-time effectively cures infrared divergences in field theories.

Changing the dimensionality of the space-time at large distances may have some consequences for cosmology. This can happen if the length scale at which the fourth spatial dimension opens up is smaller than the present cosmological horizon. Available observational data indicate that our universe is going through a phase of accelerated expansion. To date it remains a mystery what is the driving force behind the acceleration. Data favor an equation of state of the cosmic fluid $p=-\rho$, corresponding to a constant energy density. The null hypothesis is that we are observing the action of the vacuum energy density (or cosmological constant). If it is indeed the cosmological constant we are seeing, it may represent the worst prediction ever made by a theory. Instead, the cosmological constant may just be a shadow that extra dimensions cast on our visible universe. To demonstrate this, we write down the metric of  $4+1$ dimensional  space-time (with spatially isotropic $3$ dimensional slices) as
\be
ds^2 = e^{\nu }dt^2-e^{\omega }(dr^2+r^2d\Omega^2)-e^\mu d\psi^2 \, ,
\ee
where $\psi$ is the fourth spatial dimension, $d\Omega^2 \equiv d\theta^2+\sin^2\theta d\phi^2$, while  parameters $\nu, \mu$ and $\omega$ are arbitrary functions of $t$ and $\psi$. We can write down $4+1$ dimensional vacuum Einstein's equations
\be \label{GAB}
G_{AB}\equiv R_{AB}-\frac{1}{2}g_{AB}R =0
\ee
where indices $A$ and $B$ go over all of the five coordinates.
One of the homogeneous and isotropic solutions of these vacuum Einstein's equations found in \cite{PonceDeLeon:1988rg} (see also \cite{Overduin:1998pn} for a review) deserves special attention:
\be
ds^2 = dt^2 - e^{2\sqrt{\Lambda/3}\, t} \left(dr^2+r^2d\Omega^2 \right) - d\psi^2
\ee
where $\Lambda = 3/\psi^2$. This metric reduces on $\psi =$ constant hypersurfaces to a $3+1$ dimensional de Sitter metric with $\Lambda =$ constant. An observer located on a $\psi =$ constant slice of the space-time will measure the effective stress energy tensor
\be
8\pi G T_{\mu\nu} = G_{\mu\nu} \, ,
\ee
where the Greek indices go over $3+1$ dimensional hypersurface only. Note that these equations are already contained in $G_{AB} =0$. Thus, the "matter" described by $T_{\mu\nu}$ is a manifestation of pure geometry in the higher dimensional space-time and it was called "induced matter" in  \cite{Overduin:1998pn} or "shadow matter" in \cite{Frolov:2003mc,Frolov:2004bq}. The equation of state of matter defined with this stress energy tensor is $p=-\rho$ with $\rho = \Lambda/(8\pi G)$. This solution would not be of much use in theories with compact \cite{add,Starkman:2001xu,Starkman:2000dy} or warped \cite{rs} extra dimensions where the effective size of extra dimensions is small, or brane world models with non-zero tension \cite{Dvali:2000hr,deRham:2007xp} where $T_{\mu \nu}$ would get extra contributions. However, in our framework, where the three-dimensional sheet we are located on is embedded into a large structure which is effectively $4+1$ dimensional on cosmological distances, full advantage of this solution can be used. In this framework, we can address the question of the smallness of the observed cosmological constant in a completely different way. We see that observers located at different slices of five-dimensional space-time infer different values of the effective cosmological constant. The small value of the cosmological constant that we observe can be attributed to the position of our $3+1$ dimensional slice in the full $4+1$ dimensional bulk. Along these lines one might argue that small values of $\Lambda$ (i.e. large values of $\psi$) are much more natural than the opposite. Indeed, the vacuum energy density of $\rho = (10^{-3}$eV)$^4$ corresponds to
the numerical value of $\psi \sim 10^{60} M_{Pl}^{-1}$. This is comparable to the current horizon size, which is in this scenario  comparable to the characteristic distance between $3+1$ dimensional sheets comprising a $4+1$ dimensional structure we live in.

What makes this proposal of evolving dimensions very attractive is that some evidence of the lower dimensional structure of our space-time at a TeV scale may already exist. Namely,
alignment of the main energy fluxes in a target (transverse) plane has been observed in families of cosmic ray particles \cite{et1986,Mukhamedshin:2005nr,Antoni:2005ce}. The fraction of events with alignment is statistically significant for families with energies higher than TeV and large number of hadrons. This can be interpreted as evidence for coplanar scattering of secondary hadrons produced in the early stages of the atmospheric cascade development.

An interesting side-effect of such a dimensional reduction scheme is the distinct nature of gravity in lower dimensions.  It is well-known that, in a $(2+1)$-dimensional Einstein's universe, there are no local gravitational degrees of freedom, and hence there are no gravitational waves (or gravitons).  If the universe was indeed $(2+1)$-dimensional at some earlier epoch, it is reasonable to deduce that no primordial gravitational waves (PGWs) of this era exist today.  There is thus a maximum frequency for PGWs, implicitly related to the dimensional transition scale $\Lambda_2$, beyond which no waves can exist.  This indicates that gravitational wave astronomy can be used as a tool for probing the novel ``vanishing dimensions'' framework \cite{Mureika:2011bv}.

{\it Conclusions.}~
We proposed here the concept of evolving dimensions. In this concept, the space-time we live in does not have a fixed number of dimensions, but the effective number of dimensions we observe depends on the energy/lenght scale we are probing. At the length scales we are used to in everyday life and most of the observations the space is three-dimensional. The standard lore was to make our theories more complicated at short distances by introducing extra dimensions, hoping that the more complicated structure would help us resolve some fundamental problems. Instead, we could try making our theories less complicated. Indeed, it may happen that at short distances the space is lower dimensional. This will remove ultraviolet divergencies from the field theories and make the problems with the hierarchy of scales much less acute. Also, the problem of quantizing gravity becomes much simpler. Finally one can make definite predictions for the collider experiments.

In a dynamical picture where the universe starts from zero size and then grows, there is no background with the fixed number of dimensions in which the universe expands. The expanding structure actually builds dimensions during its evolution. Along these lines, it is conceivable that, at very large scales, the space becomes  effectively higher dimensional. This would make infrared divergences in field theories simply disappear. It would be interesting if our universe becomes effectively higher dimensional at scales somewhat shorter than the current horizon (very recently interesting evidence for something like that was presented in \cite{Afshordi:2008rd}). This could offer a new approach to some cosmological problems, in particular the cosmological constant problem.

A dynamical model with such an evolving background can be formulated (see e.g. \cite{Karliner:1988hd}). A string whose each segment has non-vanishing longitudinal and transverse momentum would grow in length and also cover the transverse space.

What we did not explain is the hierarchy in length scales.
This hierarchy would possibly come from the fundamental tension of the string (one-dimensional space we live on). Larger tension implies longer scale at which the string curves and/or intersects itself making the space appears two-dimensional. The transition from two- to three-dimensional space would then depend on the effective tension of the effectively two-dimensional sheet that curves and/or intersects itself making the space appears three-dimensional at larger scales. And so on. The concept of evolving dimensions is a concept of the global dynamical structure on which our mathematical theories like general relativity and field theory are formulated. Its details can vary. Here we presented some interesting consequences which are by no means exhaustive \cite{Carlip:2012md,He:2013ub,He:2011jv,Rinaldi:2012dh,Mureika:2012na,GarciaAspeitia:2011de,Nicolini:2011nz,Calcagni:2011sz,Modesto:2011kw,Obukhov:2011ks,Mann:2011rh,Nieves:2011fy,Mureika:2011py,Landsberg:2010zz,
Stoica:2013wx,GonzalezMestres:2010pi,Calmet:2010vp,Caravelli:2010be}.

{99}


\begin{thebibliography}{99}
\bibitem{Karliner:1988hd}
  M.~Karliner, I.~R.~Klebanov and L.~Susskind,
  Int.\ J.\ Mod.\ Phys.\  A {\bf 3}, 1981 (1988);
  I.~R.~Klebanov and L.~Susskind,
  Nucl.\ Phys.\  B {\bf 309}, 175 (1988).

\bibitem{Anchordoqui:2010hi}
  L.~A.~Anchordoqui, D.~C.~Dai, H.~Goldberg, G.~Landsberg, G.~Shaughnessy, D.~Stojkovic and T.~J.~Weiler,
  Phys.\ Rev.\ D {\bf 83}, 114046 (2011)  [arXiv:1012.1870 [hep-ph]].  


\bibitem{Anchordoqui:2010er}
  L.~Anchordoqui, D.~C.~Dai, M.~Fairbairn, G.~Landsberg and D.~Stojkovic,
  Mod.\ Phys.\ Lett.\ A {\bf 27}, 1250021 (2012)
  [arXiv:1003.5914 [hep-ph]].

\bibitem{Carlip:1995zj}
  S.~Carlip,
  J.\ Korean Phys.\ Soc.\  {\bf 28}, S447 (1995)
  [arXiv:gr-qc/9503024].



\bibitem{Klosch:1997md}
  T.~Klosch and T.~Strobl,
  Class.\ Quant.\ Grav.\  {\bf 14}, 1689 (1997)
  [arXiv:hep-th/9607226].

\bibitem{LouisMartinez:1993cc}
  D.~Louis-Martinez and G.~Kunstatter,
  Phys.\ Rev.\  D {\bf 49}, 5227 (1994).


\bibitem{Grumiller:2006ja}
  D.~Grumiller and R.~Meyer,
  Class.\ Quant.\ Grav.\  {\bf 23}, 6435 (2006)
  [arXiv:hep-th/0607030].

\bibitem{Zaslavskii:2003eu}
  O.~B.~Zaslavskii,
  Class.\ Quant.\ Grav.\  {\bf 20}, 2963 (2003)
  [arXiv:hep-th/0305199].


\bibitem{Giddings:1992ae}
  S.~B.~Giddings and A.~Strominger,
  Phys.\ Rev.\  D {\bf 47}, 2454 (1993)
  [arXiv:hep-th/9207034].




\bibitem{Callan:1992rs}
  C.~G.~.~Callan, S.~B.~Giddings, J.~A.~Harvey and A.~Strominger,
  Phys.\ Rev.\  D {\bf 45}, 1005 (1992)
  [arXiv:hep-th/9111056].


\bibitem{Bogojevic:1998ma}
  A.~Bogojevic and D.~Stojkovic,
  Phys.\ Rev.\  D {\bf 61} (2000) 084011
  [arXiv:gr-qc/9804070].



\bibitem{Li:1994et}
  M.~Li and C.~I.~Tan,
  Phys.\ Rev.\  D {\bf 50}, 1140 (1994)
  [arXiv:hep-th/9401134].



\bibitem{Engelhardt:1995qm}
  M.~Engelhardt and B.~Schreiber,
  Z.\ Phys.\  A {\bf 351}, 71 (1995).


\bibitem{PS} M.l E. Peskin and D. V. Schroeder, ``An Introduction to Quantum Field Theory", Addison-Wesley Advanced Book Program (1995)


\bibitem{PonceDeLeon:1988rg}
  J.~Ponce De Leon,
  Gen.\ Rel.\ Grav.\  {\bf 20}, 539 (1988).



\bibitem{Overduin:1998pn}
  J.~M.~Overduin and P.~S.~Wesson,
  Phys.\ Rept.\  {\bf 283}, 303 (1997)
  [arXiv:gr-qc/9805018].

\bibitem{Frolov:2003mc}
  V.~P.~Frolov, M.~Snajdr and D.~Stojkovic,
  Phys.\ Rev.\  D {\bf 68} (2003) 044002
  [arXiv:gr-qc/0304083].

\bibitem{Frolov:2004bq}
  V.~P.~Frolov, D.~V.~Fursaev and D.~Stojkovic,
  Class.\ Quant.\ Grav.\  {\bf 21}, 3483 (2004)
  [gr-qc/0403054].

\bibitem{add} N. Arkani-Hamed, S. Dimopoulos and G. R. Dvali,
Phys. Lett. B429, 263 (1998)

\bibitem{rs} L. Randall and R. Sundrum,
 Phys. Rev. Lett. 83, 4690 (1999)


\bibitem{Starkman:2001xu}
  G.~D.~Starkman, D.~Stojkovic and M.~Trodden,
  Phys.\ Rev.\ Lett.\  {\bf 87}, 231303 (2001)
  [hep-th/0106143].

\bibitem{Starkman:2000dy}
  G.~D.~Starkman, D.~Stojkovic and M.~Trodden,
  Phys.\ Rev.\ D {\bf 63}, 103511 (2001)
  [hep-th/0012226].



\bibitem{Dvali:2000hr}
  G.~R.~Dvali, G.~Gabadadze and M.~Porrati,
  Phys.\ Lett.\  B {\bf 485}, 208 (2000)
  [arXiv:hep-th/0005016].

\bibitem{deRham:2007xp}
  C.~de Rham, G.~Dvali, S.~Hofmann, J.~Khoury, O.~Pujolas, M.~Redi and A.~J.~Tolley,
  Phys.\ Rev.\ Lett.\  {\bf 100}, 251603 (2008)
  [arXiv:0711.2072 [hep-th]].

\bibitem{et1986} L. T. Baradzei {\it et al.} [Pamir Collaboration],  Bull.\ Russ.\ Acad.\ Sci.\ Phys.\  {\bf 50N11}, 46 (1986)
  [Izv.\ Ross.\ Akad.\ Nauk Ser.\ Fiz.\  {\bf 50}, 2125 (1986)]
\bibitem{Mukhamedshin:2005nr} R.~A.~Mukhamedshin,  JHEP {\bf 0505}, 049 (2005).

\bibitem{Antoni:2005ce}  T.~Antoni {\it et al.}  [KASCADE Collaboration],  Phys.\ Rev.\  D {\bf 71}, 072002 (2005)  [arXiv:hep-ph/0503218].



\bibitem{Mureika:2011bv}
  J.~R.~Mureika and D.~Stojkovic,
  Phys.\ Rev.\ Lett.\  {\bf 106}, 101101 (2011)  [arXiv:1102.3434 [gr-qc]].  
  J.~Mureika and D.~Stojkovic,
  Phys.\ Rev.\ Lett.\  {\bf 107}, 169002 (2011)  [arXiv:1109.3506 [gr-qc]].  




\bibitem{Afshordi:2008rd}
  N.~Afshordi, G.~Geshnizjani and J.~Khoury,
  arXiv:0812.2244 [astro-ph].


\bibitem{Carlip:2012md}
  S.~Carlip,
  AIP Conf.\ Proc.\  {\bf 1483}, 63 (2012)
  [arXiv:1207.4503 [gr-qc]].

\bibitem{He:2013ub}
  H.~-J.~He and Z.~-Z.~Xianyu,
  Phys.\ Lett.\ B {\bf 720}, 142 (2013)
  [arXiv:1301.4570 [hep-ph]].


\bibitem{He:2011jv}
  H.~-J.~He and Z.~-Z.~Xianyu,
  Eur.\  Phys.\  J.\  Plus {\bf 128}, 40 (2013)
  [arXiv:1112.1028 [hep-ph]].
  
\bibitem{Rinaldi:2012dh}
  M.~Rinaldi,
  Mod.\ Phys.\ Lett.\ A {\bf 27}, 1230008 (2012)
  [arXiv:1201.4543 [gr-qc]].
  
\bibitem{Mureika:2012na}
  J.~R.~Mureika,
  Phys.\ Lett.\ B {\bf 716}, 171 (2012)
  [arXiv:1204.3619 [gr-qc]].
  
   
\bibitem{GarciaAspeitia:2011de}
  M.~A.~Garcia-Aspeitia,
  arXiv:1109.5127 [gr-qc].
  
\bibitem{Nicolini:2011nz}
  P.~Nicolini and E.~Winstanley,
  JHEP {\bf 1111}, 075 (2011)
  [arXiv:1108.4419 [hep-ph]].
  
\bibitem{Calcagni:2011sz}
  G.~Calcagni,
  JHEP {\bf 1201}, 065 (2012)
  [arXiv:1107.5041 [hep-th]].
  
  
\bibitem{Modesto:2011kw}
  L.~Modesto,
  Phys.\ Rev.\ D {\bf 86}, 044005 (2012)
  [arXiv:1107.2403 [hep-th]].
  
\bibitem{Obukhov:2011ks}
  Y.~N.~Obukhov, A.~J.~Silenko and O.~V.~Teryaev,
  Phys.\ Rev.\ D {\bf 84}, 024025 (2011)
  [arXiv:1106.0173 [hep-th]].
  
\bibitem{Mann:2011rh}
  R.~B.~Mann and J.~R.~Mureika,
  Phys.\ Lett.\ B {\bf 703}, 167 (2011)
  [arXiv:1105.5925 [hep-th]].
  
\bibitem{Nieves:2011fy}
  J.~F.~Nieves,
  Int.\ J.\ Mod.\ Phys.\ A {\bf 26}, 5387 (2011)
  [arXiv:1105.2546 [hep-ph]].
  
\bibitem{Mureika:2011py}
  J.~R.~Mureika and P.~Nicolini,
  Phys.\ Rev.\ D {\bf 84}, 044020 (2011)
  [arXiv:1104.4120 [gr-qc]].
  
\bibitem{Landsberg:2010zz}
  G.~Landsberg,
  PoS ICHEP {\bf 2010}, 399 (2010).
  
\bibitem{Stoica:2013wx}
  O.~C.~Stoica,
  arXiv:1301.2231 [gr-qc].
  
\bibitem{GonzalezMestres:2010pi}
  L.~Gonzalez-Mestres,
  arXiv:1009.1853 [astro-ph.HE].
  
\bibitem{Calmet:2010vp}
  X.~Calmet and G.~Landsberg,
  chapter 7 in A.J. Bauer and D.G.Eiffel editors,Black Holes: Evolution, Theory and Thermodynamics Nova Publishers, New York, 2012
  [arXiv:1008.3390 [hep-ph]].
  
  
\bibitem{Caravelli:2010be}
  F.~Caravelli and L.~Modesto,
  Phys.\ Lett.\ B {\bf 702}, 307 (2011)
  [arXiv:1001.4364 [gr-qc]].



\end{thebibliography}
\end{document}